\documentstyle[axodraw,preprint,prd,aps]{revtex}

\def \to {\rightarrow}

\begin{document}
\preprint{\ \vbox{
\halign{&##\hfil\cr
       AS-ITP-2001-0012\cr\cr}} } \vfil
\draft
\title{Reexamining radiative decays of $1^{--}$ quarkonium into $\eta'$ and
$\eta$  }
\author{J.P. Ma}
\address{Institute of Theoretical Physics,\\
Academia Sinica, \\
P.O.Box 2735, Beijing 100080, China\\
e-mail: majp@itp.ac.cn}
\maketitle

\begin{abstract}
Recently CLEO has studied the radiative decay of $\Upsilon$ into $\eta'$ and
an upper limit for the decay has been determined. Confronting with this upper limit,
most of theoretical predictions for the decay fails. After briefly reviewing
these predictions we re-examine the decay by separating nonperturbative effect
related to the quarkonium and that related to $\eta'$ or $\eta$, in which the
later is parameterized by distribution amplitudes of gluons in $\eta'$. With
this factorization approach we obtain theoretical predictions which are in
agreement with experiment. Uncertainties in our predictions are discussed.
The possibly largest uncertainties are from relativistic corrections
for $J/\Psi$ and the value of the charm quark mass. We argue that the effect of
these uncertainties can be reduced by using quarkonium masses instead of using
quark masses. An example of the reduction is shown with an attempt to
explain the violation of the famous $14\%$
rule in radiative decays of charmonia.


\end{abstract}

\vfill\eject\pagestyle{plain}\setcounter{page}{1}

The gluon content of $\eta$ and $\eta'$ has been studied extensively in the
literature. For example, recent works on the subject can be found in \cite{Ball,Feld}.
Radiative decays of $1^{--}$ quarkonium into $\eta(\eta')$ provide an ideal place
to study this subject, because the decays are mediated by gluons and there is no
complication of interactions between light hadrons. Recently, CLEO has studied
the decay $\Upsilon\to \gamma +\eta'$ and an upper limit is determined\cite{CLEO}:
\begin{equation}
        {\rm Br} (\Upsilon\to\gamma+\eta') < 1.6\times 10^{-5}
\end{equation}
at $90\%$ C.L.. With this result most of theoretical predictions deliver
a branching ratio which is too large.
\par
The radiative decay has been studied in different approaches. In \cite{NR}
both the quarkonium and $\eta(\eta')$ are taken to be nonrelativistic
two-body systems, wave-functions for these bound systems are introduced.
The obtained branching ratio in this approach with a recent
compilation of $\alpha_s$ is $5-10\times 10^{-5}$\cite{NR,CLEO}
and is significantly
larger than the upper limit. In \cite{Zhao} possible mixing between
$\eta(\eta')$ and $\eta_b$ is assumed to be responsible for the decay,
the branching ratio is obtained as $6\times 10^{-5}$, which is also
larger than the upper limit. In this approach it is possible to obtain
${\rm Br} (\Upsilon\to\gamma+\eta')\approx (1\sim 3)\times 10^{-5}$
close the upper bound\cite{ZhaoP}.
\par
The corresponding decay of $J/\Psi$ has been studied by saturating a
suitable sum rule with $J/\Psi$ resonance and it has been shown
that the decay is controlled by the $U_A(1)$ anormaly\cite{Nov}.
The result of this study can be re-written in the form:
\begin{equation}
\Gamma (J/\Psi\to\gamma+\eta')= \frac{2^{11}\pi\alpha^3}{5^2\cdot 3^{12}}
                 \cdot (1-\frac{m_{\eta'}^2}{4m_c^2})^3
                 \cdot \frac{1}{m_c^4} \vert \langle 0 \vert
                 \frac{3\alpha_s}{4\pi} G^a_{\mu\nu}
                 \tilde G^{a,\mu\nu} \vert \eta'\rangle \vert^2
                 \Gamma^{-1}(J/\Psi \to e^+e^-),
\end{equation}
where $G^{a,\mu\nu}$ is the field strength tensor of gluon and $\tilde G^{a,\mu\nu}
=\frac{1}{2} \varepsilon^{\mu\nu\alpha\beta}G^a_{\ \alpha\beta}$. In the above
result we have neglected the binding energy of $J/\Psi$ and taken
$M_{J/\Psi}=2m_c$, $m_c$ is the pole mass of c-quark. If one can
generalize the approach for the $\Upsilon$ decays, one can obtain
the ratio:
\begin{equation}
R_{\eta'} = \frac {{\rm Br}(\Upsilon\to\gamma+\eta')}
   {{\rm Br}(J/\Psi\to\gamma+\eta')}
      =\left [ \frac {\Gamma(J/\psi\to e^+e^-)}
   {\Gamma(\Upsilon\to e^+e^-)}\cdot \frac {\Gamma(J/\psi\to X)}
   {\Gamma(\Upsilon\to X)} \right ]
   \cdot \frac {m_c^4}{m_b^4}\cdot
   \frac {(1-\frac {m_{\eta'}^2}{4m_b^2})^3}{(1-\frac {m_{\eta'}^2}{4m_c^2})^3}.
\end{equation}
Using the experimental results for the widths in the bracket we obtain
\begin{equation}
R_{\eta'} \approx 6.6\cdot \frac {m_c^4}{m_b^4}\cdot
   \frac {(1-\frac {m_{\eta'}^2}{4m_b^2})^3}{(1-\frac {m_{\eta'}^2}{4m_c^2})^3}.
\end{equation}
The branching ratio of $J/\Psi$ has been measured and its value
is $(4.31\pm 0.3)\times 10^{-3}$. We take the quark masses as
$m_b=M_{\Upsilon}/2\approx 5$GeV and $m_c=M_{J/\Psi}/2\approx 1.5$GeV and obtain
${\rm Br}(\Upsilon\to\gamma+\eta')\approx 3.1\times 10^{-4}$, which
is too large for the upper limit. However, the generalization
of Eq.(2) to $\Upsilon$ may not be correct. In the spirit of the
approach the emitted gluons, which are converted into $\eta'$,
are soft, while in $\Upsilon$ decay the gluons are definitely
hard. Employing multipole expansion for the soft gluons
one is also able to predict the decay of $J/\Psi$\cite{Kuang}
\par
From above discussions one may conclude that the predictions based
on QCD-inspired models or on sum rule are not compatible
with the upper limit, or not consistent. It should be also noted
that phenomenological models can have compatible predictions.
In an extended vector-dominance model one indeed finds the branching
ratio from $5.3\times 10^{-7}$ to $2.5\times 10^{-6}$\cite{Intemann},
but this model has no direct relation to QCD as the fundamental theory
of strong interaction.
\par
Decays of quarkonia were intensively studied in eighties. Now our understanding
of QCD has been greatly improved, a restudy of these decays is necessary to
explain new experimental results like the upper bound in Eq.(1).
On the other hand, there is a large data sample with $5\times 10^7$ $J/\psi$
events collected with BES\cite{zhao},
a date sample with several billions $J/\Psi$ events
is planed to be collected with the
proposed BES III at BEPC II and with CLEO-C at modified Cornell Electron
Storage Ring (CESR)\cite{zhao,cleoc}. Furthermore, about
$4 ~{\rm fb}^{-1}$ $b{\bar b}$ resonance data are
planned to be taken at CLEO III in the year prior to conversion to low
energy operation (CLEO-C)\cite{cleoc}. These data samples of quarkonia
will allow us to study the decays, which have been observed before, with
more accuracy, and also those decays which have been not observed. Therefore,
experimental activities will bring more information about these decays
and may also lead to new discoveries, e.g., discovery of glueball.
In this brief report we present an approach based on QCD factorization
to explain the experimental result from CLEO\cite{CLEO}.
This approach was used for the radiative decay into the tensor meson
$f_2$\cite{Ma}.
\par
We consider the heavy quark limit, i.e., $m_c\to\infty,\ m_b\to\infty$.
In the limit, a quarkonium system, taking $\Upsilon$ as an example,
can be taken as a bound state of $b$- and $\bar b$-quark which move
with a small velocity $v$,
hence
an expansion in $v$ can be employed, nonrelativistic QCD(NRQCD) can be used
to describe the nonperturbative effect related to $\Upsilon$\cite{BBL}.
The decay can be thought as the following:
the quarkonium will be annihilated into a real photon and gluons, the gluons
will be subsequently converted into the meson $\eta'$.
Also in the limit,
the meson $\eta'$ has a large momentum, this enables an expansion in twist
to characterize the gluonic conversion into $\eta'$, the conversion is then described
by a set of distribution amplitudes of gluons. The large
momentum of $\eta'$ requires that the gluons should be hard, hence
the emission of the gluons can be handled by perturbative theory.
The above discussion implies that
we may factorize the decay amplitude into three parts: the first part consists
of matrix elements of NRQCD representing the nonperturbative effect related
to $\Upsilon$, the second part consists of some distribution amplitudes, which
are for the gluonic conversion into $\eta'$, the third part consists of some
coefficients, which can be calculated with perturbative theory for
the $b\bar b$-pair annihilated into gluons and a real photon.
In this report we show that the contribution of twist-2 operators are suppressed
by $m_{\eta'}^2$. This indicates that a complete QCD-analysis should include
contributions from twist-4 operators. However, without such a complete analysis
one still can make some predictions like the branching ratio given in Eq.(1).
\par
We consider the decay of $\Upsilon$
\begin{equation}
\Upsilon\to \gamma(q)+\eta'(k),
\end{equation}
where the momenta are given in the brackets. We take a light-cone coordinate
system, in which the momentum $k$ of $\eta'$ is $k^\mu =(k^+,k^-,0,0)$. We consider
the contribution from emission of two gluons, and assume a factorization
can be performed. Then the $S$-matrix can be written as:
\begin{eqnarray}
 \langle \gamma \eta' |S|\Upsilon\rangle &=&-i\frac 12e Q_bg_s^2 \varepsilon^*_\rho \int
d^4xd^4yd^4zd^4x_1d^4y_1 e^{iq\cdot z}\langle \eta' |G_\mu ^a(x)G_\nu
^b(y)|0\rangle  \nonumber \\
&&\langle 0|\bar{b}_j(x_1)b_i(y_1)|\Upsilon\rangle \cdot M_{ji}^{\mu \nu \rho
,ab}(x,y,x_1,y_1,z),
\end{eqnarray}
where $M_{ji}^{\mu \nu \rho ,ab}(x,y,x_1,y_1,z)$ is a known function, $i$
and $j$ stand for Dirac- and color indices, $a$ and $b$ is the color of
gluon field, $b(x)$ stands for the Dirac field of $b$-quark,
$\varepsilon^*$ is the polarization vector of the photon,
$Q_b$ is the charge fraction of the $b$-quark in unit $e$.
The above equation can be generalized to emission of arbitrary
number of gluons. Using the fact that $b$-or $\bar b$-quark moves
with a small velocity $v$, the matrix element with the Dirac fields can
be expanded in $v$. We obtain:
\begin{equation}
\langle 0|\bar{b}_j(x)b_i(y)|\Upsilon \rangle =-\frac 16(P_{+}\gamma ^\ell
P_{-})_{ij}\langle 0|\chi ^{\dagger }\sigma ^\ell \psi |\Upsilon \rangle
e^{-ip\cdot (x+y)}+{\cal O}(v^2),
\end{equation}
where $\chi ^{\dagger }(\psi )$ is the NRQCD field for $\bar{b}(b)$ quark
and
\begin{eqnarray}
P_{\pm } &=&(1\pm \gamma ^0)/2,  \nonumber \\
p^\mu &=&(m_b,0,0,0),
\end{eqnarray}
where $m_b$ is the pole-mass of the $b$-quark. In Eq.(7) we do not count the power of
$v$ for quark fields because this power is same for every term in the expansion
of Eq.(7). With this in mind
the leading order of the matrix element is then ${\cal O}(v^0)$, we will
neglect the contribution from higher orders and the momentum of $\Upsilon $ is
then approximated by $2p$. It should be noted that effects at higher order
of $v$ can be added with the expansion in Eq.(7).
\par
For the matrix element with gluon fields we observe that
the $x$-dependence of the matrix element
is controlled by different scales: the $x^-$-dependence is controlled
by $k^+$, while the $x^+$- and $x_T$-dependence is controlled by
the scale $\Lambda_{QCD}$ or $k^-$, which are small in comparison with $k^+$.
Because of these small scales we can expand the matrix element in $x^+$ and in
$x_T$. With this expansion we obtain the result for Fourier transformed
matrix element:
\begin{eqnarray}
\int dx^4 e^{-iq_1\cdot x}
 \langle\eta' \vert G^{a,\mu}(x) G^{b,\nu}(0) \vert 0\rangle
 &=& \frac{1}{8}\delta^{ab}
(2\pi)^4 \delta(q_1^-) \delta^2 (q_{1T})
\cdot \frac{1}{2k^+x_1(x_1-1)} e^{\mu\nu}  F_{\eta'}(x_1)  \nonumber\\
&& + \cdots,
\end{eqnarray}
with
\begin{eqnarray}
e^{\mu\nu}&=&\varepsilon^{\mu\nu\alpha\beta} l_\alpha n_\beta,
\ \ l^\mu =(1,0,0,0),\ \ n^\mu=(0,1,0,0), \ \ q_1^+=x_1k^+, \nonumber\\
F_{\eta'} (x_1)&=& \frac{1}{2\pi k^+} \int dx^-e^{-ix_1k^+x^-}
\langle \eta' (k) \vert G^{a,+\mu}(x^-) G^{a,+\nu} (0) \vert 0\rangle
 e_{\mu\nu},
\end{eqnarray}
where $\varepsilon^{\mu\nu\alpha\beta}$ is totally anti-symmetric
with $\varepsilon^{0123}=1$. $F_{\eta'} (x_1)$ is
the distribution amplitude characterizing
the conversion of two gluons into $\eta'$ and it is defined with twist-2
operators in the light-cone gauge. In other gauges a gauge link should
be supplied in Eq.(10) to maintain the gauge invariance.
It should be noted that there is no simple relation between
$F_{\eta'} (x_)$ and the gluonic matrix element in Eq. (2). The $\cdots$
in Eq.(9) stands for contributions from higher twist. The next-to-leading
twist is 4. With the above results we obtain the $S$-matrix element with
the twist-2 contribution in the limit $m_{\eta'}\to 0$:
\begin{eqnarray}
\langle \gamma \eta'  |S|\Upsilon\rangle &=&
  \frac {-i}{48}eQ_bg_s^2(2\pi )^4\delta^4(2p-k-q)\varepsilon^*_\rho
\langle 0|\chi ^{\dagger }\sigma ^\ell \psi |\Upsilon  \rangle \
  e^{\ell\rho} \frac{1}{m_b^4}
\nonumber\\
 && \cdot m^2_{\eta'}  \int dx_1 \frac{1-2x_1}{x_1(1-x_1)^2}
  F_{\eta'} (x_1)\cdot \left( 1 +
  {\cal O}(\frac {m_{\eta'}^2}{m_b^2}) \right).
\end{eqnarray}
It shows that the twist-2 contribution is suppressed by $m_{\eta'}^2$.
In the twist-expansion the light hadron mass $m_{\eta'}$ should be
taken as a small scale as $\Lambda_{QCD}$, hence the contribution
is proportional to $\Lambda^2_{QCD}$. This implies that a complete
analysis at the leading order should include not only this contribution but also
twist-4 contributions,
in which one needs to consider
the contributions from emission of 2, 3 and 4 gluons. This
is too complicated to be done here.
However, without a complete
analysis we can always write the result of a complete analysis
as
\begin{equation}
\langle \gamma \eta'  |S|\Upsilon\rangle =
  \frac {-i}{48}eQ_bg_s^2(2\pi )^4\delta^4(2p-k-q)\varepsilon^*_\rho
\langle 0|\chi ^{\dagger }\sigma ^\ell \psi |\Upsilon  \rangle \
  e^{\ell\rho} \frac{1}{m_b^4}\cdot g_{\eta'},
\end{equation}
where the parameter $g_{\eta'}$ has a dimension 3 in mass. This parameter
is a sum of the twist-2 contribution in the second line of Eq.(11) and
the twist-4 contributions which need to be analyzed. The parameter
characterized the conversion of gluons
into $\eta'$ and it does not depend on properties of $\Upsilon$.
The origin of the factor $m_b^4=m_b^2\cdot m_b^2$ is: one $m_b^2$ comes from
the perturbative part, another $m_b^2$ reflects the fact that the contribution
of twist-2 operators is proportional to $m^2_{\eta'}$ and contributions
of twist-4 are proportional to $\Lambda^2_{QCD}$. It is interesting
to note that this power behavior is also obtained in \cite{Bai}, in contrast,
it is also pointed out in \cite{Bai} that this behavior holds by taking $m_c$ and
$m_b$ as light quark masses, in the heavy quark limit this behavior does not
hold.
\par
The above result may also be generalized for the decay of $J/\Psi$.
One may have a doubt that the twist expansion may not be applicable
for the $J/\Psi$ decay, because $m_c$ is not large enough. The
twist expansion means a collinear expansion of momenta of partons
in $\eta'$, components of these momenta have the
order of $({\cal O}(k^+),{\cal O}((k^-),{\cal O}(\Lambda_{QCD}),
{\cal O}(\Lambda_{QCD}))$. Hence the expansion parameters are
\begin{equation}
\frac{k^-}{k^+}=\frac{m^2_{\eta'}}{M_{J/\Psi}^2}\approx 0.1, \ \
  \frac {\Lambda_{QCD}}{k^+}\approx  0.2,
\end{equation}
where we have taken $\Lambda_{QCD}\approx 400$MeV. These
estimations show that the twist expansion is also a good
approximation for the $J/\Psi$ decay. There are also
other possible large uncertainties, due to effects from higher
orders of $v$ and $\alpha_s$. These can be eliminated partly
by building the ratio $\Gamma (J/\Psi(\Upsilon)\to\gamma\eta')/
\Gamma(J/\Psi(\Upsilon)\to{\rm light\ hadrons)}$. Theoretical prediction
for this ratio will have less uncertainties than the width,
because corrections from higher orders of $v$ and $\alpha_s$
are cancelled at certain level. Using experimental data
for ${\rm Br}(J/\Psi(\Upsilon)\to{\rm light\ hadrons)}$ we can predict
the branching ratio. With this consideration we rewrite the ratio defined
in Eq.(3) as:
\begin{eqnarray}
R_{\eta'} &=& \frac {{\rm Br}(\Upsilon\to{\rm light\ hadrons})}
                 {{\rm Br}(J/\Psi\to{\rm light\ hadrons})} \cdot r_{\eta'},
\nonumber\\
 r_{\eta'} &=& \frac {\Gamma (\Upsilon\to\gamma\eta')/
\Gamma(\Upsilon\to{\rm light\ hadrons)}}
  {\Gamma (J/\Psi\to\gamma\eta')/
\Gamma(J/\Psi\to{\rm light\ hadrons)}}\approx \frac {Q_b^2 m_c^6}{Q_c^2m_b^6}\cdot
   \frac{\alpha_s(m_c)}{\alpha_s(m_b)},
\end{eqnarray}
where leading order results for the decay widths are used for
$r_{\eta'}$.
Using the experimental results
for the branching ratios of decays into light hadrons, we obtain:
\begin{equation}
R_{\eta'}=  \frac {{\rm Br}(\Upsilon\to\gamma+\eta')}
    {{\rm Br}(J/\Psi\to\gamma+\eta')}\approx 1.31
    \cdot \frac {Q_b^2 m_c^6}{Q_c^2m_b^6}\cdot
   \frac{\alpha_s(m_c)}{\alpha_s(m_b)}.
\end{equation}
This is the result at the leading order of $\Lambda$, where $\Lambda$ is
$\Lambda_{QCD}$ or $m_{\eta'}$, and the dependence of the renormalization
scale in gluonic distribution amplitudes is neglected. The dependence
may be extracted from the study in \cite{LBG}.
By taking
$\alpha_s(m_c)\approx 0.3$ and $\alpha_s(m_b)\approx 0.18$ we obtain:
\begin{equation}
R_{\eta'} \approx 3.9\times 10^{-4}.
\end{equation}
With the experimental value of ${\rm Br}(J/\Psi\to\gamma+\eta')$ we
obtain the branching ratio:
\begin{equation}
{\rm Br}(\Upsilon\to\gamma+\eta')\approx 1.7\times 10^{-6}.
\end{equation}
This value is much smaller than the values obtained with other
approaches and it is in consistency with the upper limit. Similarly
we also obtain
\begin{equation}
{\rm Br}(\Upsilon\to\gamma+\eta)\approx 3.3\times 10^{-7}.
\end{equation}
It should be emphasized that our results obtained in the above equations
are not based on any model, corrections to these results can be systematically
added in the framework of QCD. The possibly largest uncertainties in our results
are from relativistic corrections for $J/\Psi$ decays and from the uncertainty of
the value of the charm quark mass, each of them can be at the level of $50\%$.
Taking these into account, our prediction in Eq.(17) and (18) can be close
to the experimental bound in Eq.(1). However, these largest uncertainties may be reduced
by using hadron masses, i.e., using $2m_c=M_{J/\Psi}$. This possibility
is based on the result for relativistic correction in \cite{GK} and
on the observation that the violation of the
famous $14\%$ rule may be reduced in this way. If one analyzes the correction at the
next-to leading order of $v$ for decays of $1^{--}$ quarkonia, one obtains that the
correction is proportional to a NRQCD matrix element defined in \cite{BBL}. This matrix
element represents a relativistic correction. In \cite{GK} it is shown that this matrix
element is proportional to the binding energy, i.e., to $M_{J/\Psi}-2m_c$ for $J/\Psi$
and to $M_{\psi'}-2m_c$ for $\psi'$ respectively. If we use $2m_c=M_{J/\Psi}$ for
$J/\Psi$ decays and $2m_c=M_{\psi'}$ for $\psi'$ decays respectively, the relativistic
correction disappears formally, but it is actually included by using hadron masses. However,
it should be noted that this should be regarded as a phenomenological estimation,
a detailed analysis and an precise determination of quark masses is needed to
study the correction in a consistent way.
\par
The famous $14\%$ rule is derived simply by taking leading order results
for decays. In our case we have:
\begin{equation}
\frac {{\rm Br}(\psi'\to\gamma\eta')}{{\rm Br} (J/\Psi\to\gamma\eta')}
   =\frac {{\rm Br}(\psi'\to e^+e^-)}{{\rm Br} (J/\Psi\to e^+e^-)}=0.147\pm 0.023 ,
\end{equation}
where the number is estimated with experimental results of leptonic decay widths.
This result is theoretically expected not only
for radiative decays into any light hadron, but also for hadronic
decays, this is the so-called $14\%$ rule. However, this rule is significantly
violated, one of the violations is the well known $\rho\pi$ puzzle. A possible
explanation and useful references can be found in \cite{Chen}.
The experimental result made by BES\cite{BES} indicates that the rule is also
violated in our case:
\begin{equation}
\frac {{\rm Br}(\psi'\to\gamma\eta')}{{\rm Br} (J/\Psi\to\gamma\eta')}
 = 0.036\pm 0.009.
\end{equation}
This value is only fourth of the expected. It should be noted that
corrections from higher orders of $\alpha_s$ is cancelled in the
ratios in Eq.(18), the theoretical uncertainties come from effects
of higher orders in $v$ in Eq.(7) and higher twists. In the case
of $1^{--}$ quarkonia, the correction from the next-to-leading order
of $v$ is the relativistic correction, whose effect is expected
to be significant for charmonia.  As discussed before, this correction may be
estimated by replacing $m_c$ with the half of the mass of quarkonium,
i.e., we use $2m_c=M_{J/\Psi}$ for the $J/\Psi$ decays and
$2m_c=M_{\psi'}$ for $\psi'$ decays.
With this replacement and with our result in Eq.(12),
the ratio in Eq.(19) is modified as:
\begin{equation}
\frac {{\rm Br}(\psi'\to\gamma\eta')}{{\rm Br} (J/\Psi\to\gamma\eta')}
   =\frac {M^6_{J/\Psi}}{M^6_{\psi'}}
   \frac {{\rm Br}(\psi'\to e^+e^-)}{{\rm Br} (J/\Psi\to e^+e^-)}=0.0512\pm 0.0080,
\end{equation}
This result shows that the relativistic correction is indeed significant.
With the replacement the predicted ratio is much closer to the experimental
result than that of the $14\%$ rule and the two largest uncertainties
are reduced in the prediction.
However this is a naive estimation, a detailed study is needed and is in
progress\cite{ChenMa}. With this case we can expect that the two largest
uncertainties are also reduced in our predictions in Eq.(17) and (18)
because we have used $2m_b=M_{\Upsilon}$ and $2m_c=M_{J/\Psi}$.
\par
It is also interesting to look at decays into $\rho\pi$. In this decay
one of the final hadrons is produced at the level of twist-2, another
is at the level of twist-3\cite{CZ}. With this fact and with the
replacement the rule is modified as:
\begin{equation}
Q_{\rho\pi}=\frac {{\rm Br}(\psi'\to\rho\pi)}{{\rm Br} (J/\Psi\to\rho\pi)}
   =\frac {M^8_{J/\Psi}}{M^8_{\psi'}}
   \frac {{\rm Br}(\psi'\to e^+e^-)}{{\rm Br} (J/\Psi\to e^+e^-)}=0.036\pm 0.006.
\end{equation}
With the modification the rule is changed significantly.
The above result also holds for decays into $K^*K$. Although
the ratio is reduced, but it is still in conflict with experimental
results. In \cite{Mark} it is found that $Q_{\rho\pi}<0.006$
and $Q_{K^{*+}K^-}<0.64$. Recent data from BES gives
$Q_{\rho\pi}<0.0023$ and $Q_{K^{*0}{\bar K}^0}=0.017\pm0.006$\cite{Zhu}.
However, the predictions are closer to experimental than that in Eq.(19).
One should also keep in mind that these decays are more complicated
than radiative decays discussed before, because the final state
consists of two light hadrons.

\par

To summarize: We have presented a QCD-factorization approach
for radiative decays of $1^{--}$ quarkonium into $\eta(\eta')$,
the result is in consistency with the experimental result made
by CLEO. On the other hand, most of theoretical results
is not compatible with the upper limit. A possible explanation
for the violation of the $14\%$ in our case is given. With this
explanation we show that effect of relativistic corrections
and that due to uncertainty of the quark mass can be reduced
by using quarkonium masses and uncertainties in our predictions
may be not so large as those usually expected.

\vskip 10mm
\begin{center}
{\bf\large Acknowledgments}
\end{center}

The work is supported  by National Science Foundation
of P. R. China and by the Hundred Young Scientist Program of
Academia Sinica of P. R. China.

\end{document}